\documentclass[11pt,twoside]{article}

\usepackage{asp2014}

\aspSuppressVolSlug
\resetcounters

\begin{document}

\title{Provenance as a requirement for large-scale complex astronomical instruments}


\author{Mathieu Servillat,$^1$ Catherine Boisson,$^1$ Julien Lefaucheur,$^1$ Karl Kosack,$^2$ Mich\`{e}le Sanguillon,$^3$ Mireille Louys,$^{4,5}$ and Fran\c{c}ois Bonnarel$^4$ \\
\affil{$^1$Laboratoire Univers et Th\'{e}ories, Observatoire de Paris, PSL Research University, CNRS, 92190 Meudon, France; \email{mathieu.servillat@obspm.fr}}
\affil{$^2$CEA Saclay, DSM/IRFU/SAp, Bat 709, F-91191 Gif-Sur-Yvette, France}
\affil{$^3$Laboratoire Univers et Particules de Montpellier, Universit\'{e} de Montpellier, CNRS/IN2P3, France}
\affil{$^4$Centre de Donn\'{e}es astronomiques de Strasbourg, Observatoire Astronomique de Strasbourg, Universit\'{e} de Strasbourg, CNRS, Strasbourg, France}
\affil{$^5$ICube Laboratory, Universit\'{e} de Strasbourg, CNRS, Strasbourg, France}}

\paperauthor{Mathieu Servillat}{mathieu.servillat@obspm.fr}{0000-0001-5443-4128}{Observatoire de Paris, PSL Research University, CNRS}{LUTH}{Meudon}{}{92195}{France}
\paperauthor{Catherine Boisson}{catherine.boisson@obspm.fr}{}{Observatoire de Paris, PSL Research University, CNRS}{LUTH}{Meudon}{}{92195}{France}
\paperauthor{Julien Lefaucheur}{julien.lefaucheur@obspm.fr}{}{Observatoire de Paris, PSL Research University, CNRS}{LUTH}{Meudon}{}{92195}{France}
\paperauthor{Karl Kosack}{karl.kosack@cea.fr}{}{CEA Saclay}{DSM/IRFU/SAp}{Gif-Sur-Yvette}{}{91191}{France}
\paperauthor{Mich\`{e}le Sanguillon}{@email.edu}{0000-0003-0196-6301}{Universit\'{e} de Montpellier, CNRS}{LUPM}{Montpellier}{}{34095}{France}
\paperauthor{Mireille Louys}{mireille.louys@unistra.fr}{0000-0002-4334-1142}{Universit\'{e} de Strasbourg, CNRS} {ICube Laboratory - UMR7357}{Strabourg}{}{67000}{France}
\paperauthor{Fran\c{c}ois Bonnarel}{francois.bonnarel@astro.unistra.fr}{}{Universit\'{e} de Strasbourg, CNRS}{Observatoire astronomique de Strasbourg - UMR7550}{Strabourg}{}{67000}{France}

\begin{abstract}

We developed several pieces of software to enable the tracking of provenance information for the large-scale complex astronomical observatory CTA, the Cherenkov Telescope Array. Such major facilities produce data that will be publicly released to a large community of scientists. There are thus strong requirements to ensure data quality, reliability and trustworthiness. Among those requirements, traceability and reproducibility of the data products have to be included in the development of large projects. Those requirements can be answered by structuring and storing the provenance information for each data product.

We followed the Provenance data model, currently discussed at the IVOA, and implemented solutions to collect provenance information during the CTA data processing and the execution of jobs on a work cluster.

\end{abstract}


\section{Introduction}

State of the art observations are now performed by large-scale complex astronomical instruments. A consortium of specialists is generally responsible for the development and the operation of large observatories, as it is the case for example for the Cherenkov Telescope Array\footnote{\url{http://www.cta-observatory.org/}} (CTA). The path of the data production from acquisition to dissemination, through e.g. data centers, archives and web portals, can be extremely obscure to the end user. This complexity is illustrated in Figure~\ref{P4-75_f1}.

\articlefigure[width=\textwidth]{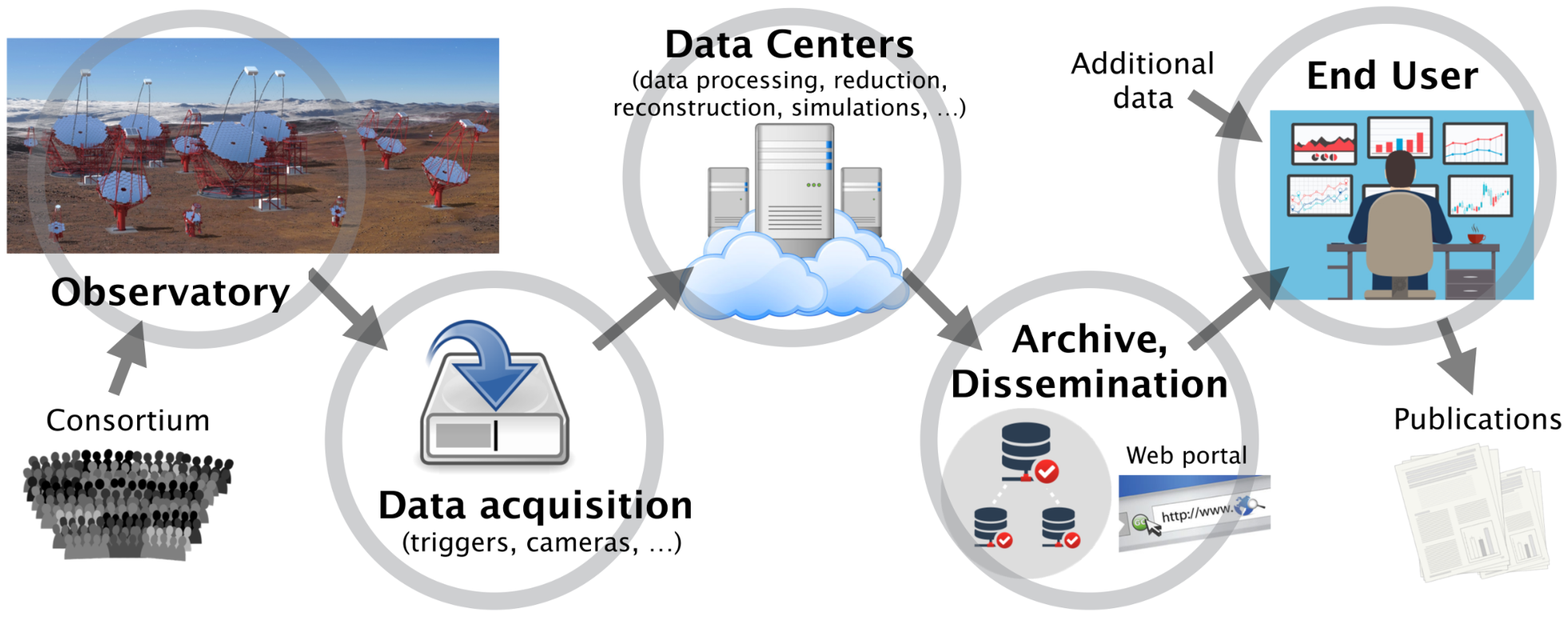}{P4-75_f1}{Data production for large-scale complex astronomical instruments such as CTA.}

In order to assess the usefulness and the quality of the data for their own scientific work, end users need a flowchart explaining the large number of steps and complexity involved in the data preparation. This can be done by collecting provenance information at each step of the data preparation. We followed the IVOA Provenance data model \citep{IVOAProvenanceDM,P129_adassxxvii} to develop solutions for CTA. 

Provenance is information about entities, activities, and people involved in producing a piece of data\citep{std:W3CProvDM}. It helps to trace back the data lineage through the production pipeline, and learn about the methods used and the people or organizations involved in the project.

\section{How to collect provenance information during CTA data production?}

The production of scientific data from CTA will use a complex and specific Pipeline, accessing different resources and calibration products, and using complex algorithms. A key feature of the Pipeline that was included early in the development is the storage of provenance information at each step of the data processing.

In order to enable the recording of Provenance information, the development followed those steps, where it was important to include the notions of Provenance early in the data model design: 

\begin{itemize}
\item Include the relevant metadata in the CTA data model \citep{servillat_adassxxvi}
\item Follow the IVOA Provenance data model for the generated data \citep{riebe_adassxxvi}
\item Collect provenance information at each step of the data processing:
\begin{itemize}
\item Use unique identifiers for entities, activities and agents
\item Describe each task executed in the Pipeline
\item Keep a list of all used and generated entities during the execution of an activity
\end{itemize}
\end{itemize}

A Provenance Python class has been developed for the CTA Pipeline framework \texttt{ctapipe}\footnote{\url{https://github.com/cta-observatory/ctapipe}}. This class is loaded automatically when a task is executed and provenance information is automatically recorded : when the task is started, when it ends, when an input entity (file, database access) is touched and when an output entity is created. 

This makes the collection of provenance information mostly hidden to the user, but also to the developers. The resulting dictionary at the end of the task could be combined with a description of the task to generate an IVOA Provenance compatible file, adding in particular links to persons responsible for the task.

The Provenance class serves different goals, first the tracking of the history of a data product to inform the end user about its origin and quality, but also the possibility to check the integrity of the Pipeline an locate sources of errors by searching structured provenance information.

\section{How to store and expose the provenance information in a standard format?}

We developed a job control system that stores provenance information following the IVOA UWS pattern and Provenance data model. 
OPUS\footnote{\url{https://github.com/mservillat/OPUS}} (Observatoire de Paris UWS System) is a light job control system developed as an open source Python application.

The following features have been implemented:

\begin{itemize}
\item Edit and fill Activity Descriptions, following the proposed IVOA standard serialization \citep{IVOAProvenanceDM}
\item Run jobs asynchronously on a work cluster. OPUS connects with the workload manager used at the Observatoire de Paris (SLURM -- Simple Linux Utility for Resource Management), but it can also run jobs on the local computer/server.
\item Present the list of jobs attached to a user per available job.
\item Present a status page for each job with input and results.
\item Generate and return Provenance files after job completion, that are attached to the job as results.
\end{itemize}

This system has been used to test the execution of CTA data analysis tools on a work cluster, as it can be seen in Figure~\ref{P4-75_f2}. Such a service can be included in a data access web portal as it is currently tested in the CTA Data Distiller prototype\footnote{\url{https://voparis-cta-test.obspm.fr}}.

\articlefigure[width=\textwidth]{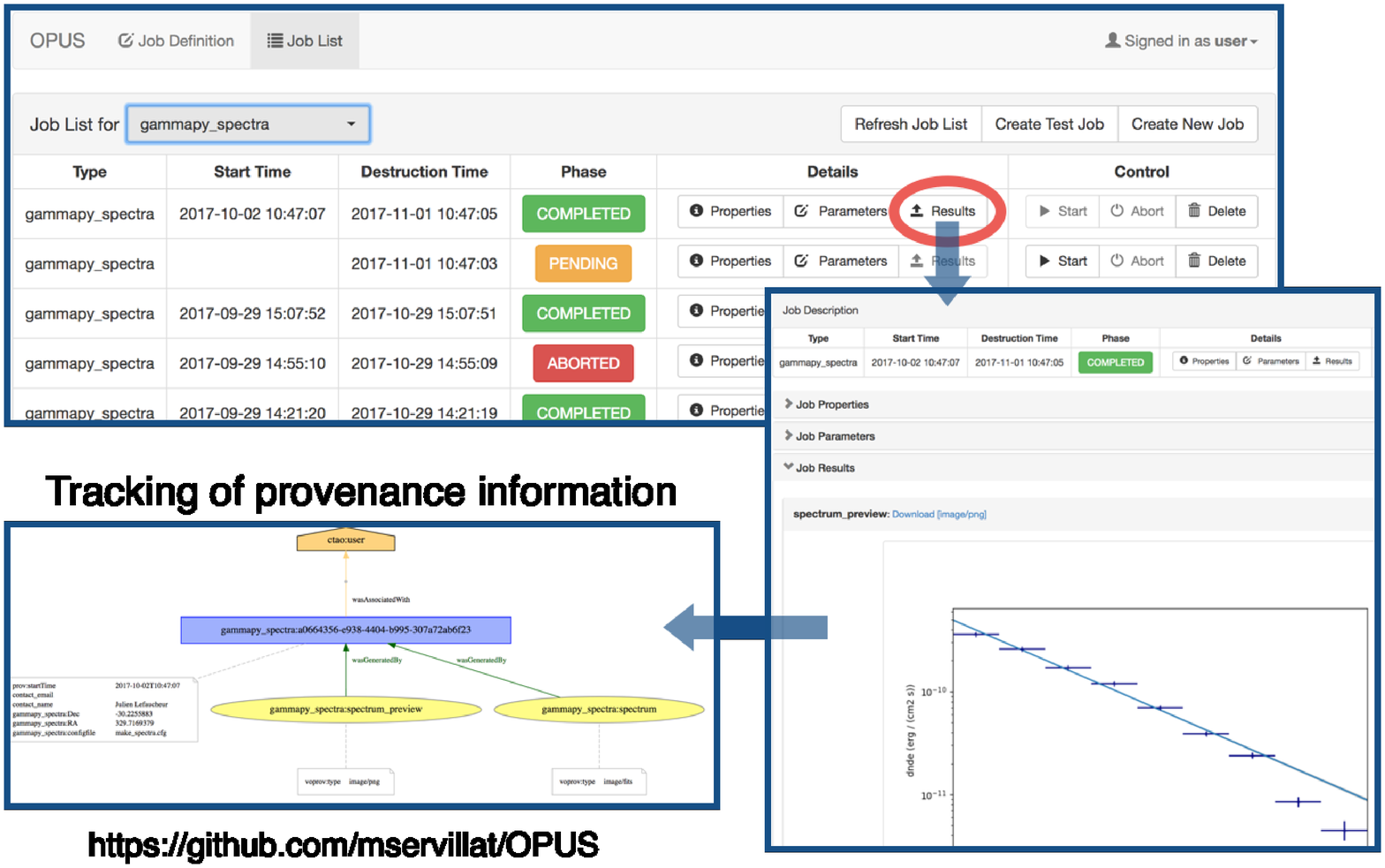}{P4-75_f2}{Screenshots of OPUS for a CTA related job. The joblist for the user and the job \emph{gammapy\_spectra} is shown at the top, then the result page with preview of the spectra is shown pn the right, and the provenance tree for this job is attached in PROV format on the left.}

\section{Conclusion}

We developed tools that implement the IVOA Provenance proposed standard in the context of a large-scale complex astronomical observatory, with the aim to provide generic tools that can be used for other projects.

\acknowledgements This work was partially funded by ASTERICS (\url{http://www.asterics2020.eu/}), a project supported by the European Commission Framework Programme Horizon 2020 Research and Innovation action under grant agreement n. 653477; Additional funding was provided by the INSU (Action Sp\'ecifique Observatoire Virtuel, ASOV), the Action F\'ed\'eratrice CTA at the Observatoire de Paris and the Paris Astronomical Data Centre (PADC).

\bibliography{P4-75}  

\end{document}